# Periodic Emission from the Gamma-ray Binary 1FGL J1018.6−5856*

The Fermi-LAT Collaboration†

**Gamma-ray binaries are stellar systems containing a neutron star or black hole with gamma-ray emission produced by an interaction between the components. These systems are rare, even though binary evolution models predict dozens in our Galaxy. A search for gamma-ray binaries with the Fermi Large Area Telescope (LAT) shows that 1FGL J1018.6−5856 exhibits intensity and spectral modulation with a 16.6 day period. We identified a variable X-ray counterpart, which shows a sharp maximum coinciding with maximum gamma-ray emission, as well as an O6V((f)) star optical counterpart and a radio counterpart that is also apparently modulated on the orbital period. 1FGL J1018.6−5856 is thus a gamma-ray binary, and its detection suggests the presence of other fainter binaries in the Galaxy.**

Two types of interacting binaries containing compact objects are expected to emit gamma rays (*1*): microquasars – accreting black holes or neutron stars with relativistic jets (*2*) – and rotation-powered pulsars interacting with the wind of a binary companion (*3*). Microquasars should typically be powerful X-ray sources when active, and hence such gamma-ray emitting systems may already be known X-ray binaries. Indeed, the bright X-ray source Cygnus X-3 is now known to be such a source (*4, 5*). The existence of pulsars interacting with early spectral type stellar companions is predicted as an initial stage in the formation of high-mass X-ray binaries (HMXBs) containing neutron stars (*6*). These interacting pulsars are predicted to be much weaker X-ray emitters, and may not yet be known, or classified, X-ray sources. Gamma-ray binaries may thus not be as rare as they appear to be, and many systems may await detection.

†All authors with their affiliations appear at the end of this paper.

A gamma-ray binary is expected to show orbitally-modulated gamma-ray emission due to a combination of effects, including changes in viewing angle and, in eccentric orbits, the degree of the binary interaction, both of which depend on binary phase. Periodic gamma-ray modulation has indeed been seen in LS 5039 (3.9 day period), LS I +61° 303 (26.5 days), and Cygnus X-3 (4.8 hours) (*4, 7, 8*), and gamma-ray emission is at least orbital phase dependent for the PSR B1259−63 system (3.4 years) (*9*). However, the putative gamma-ray binary HESS J0632+057, for which a 321 day X-ray period is seen, has not yet been shown to exhibit periodic gamma-ray emission (*10*). PSR B1259−63 contains a pulsar, and LS 5039 and LS I +61° 303 are suspected, but not proved, to contain pulsars, whereas Cygnus X-3 is a black hole candidate. A search for periodic modulation of gamma-ray flux from LAT sources may thus lead to the detection of further gamma-ray binaries, potentially revealing the predicted HMXB precursor population. The first Fermi LAT (*11*) catalog of gamma-ray sources ("1FGL") contains 1451 sources (*12*), a large fraction of which do not have confirmed counterparts at other wavelengths and thus are potentially gamma-ray binaries.

In order to search for modulation we generated light curves for all 1FGL sources in the energy range 0.1 – 200 GeV employing a weighted photon method (see Supporting Online Material; SOM). We then calculated power spectra for all sources. From an examination of these, in addition to modulation from the known binaries LS I +61° 303 and LS 5039, we noted the presence of a strong signal near a period of 16.6 days from 1FGL J1018.6−5856 (Fig. 1). 1FGL J1018.6−5856 has a cataloged 1 – 100 GeV flux of 2.9×10$^{-8}$ photons cm$^{-2}$ s$^{-1}$, making it one of the brighter LAT sources. The source's location at right ascension (R.A.) = 10$^h$ 18.7$^m$, declination (decl.) = −58° 56.3′ (J2000; ± 1.8′, 95% uncertainty) means that it lies close to the Galactic plane ($b$ = $-1.7°$), marking it as a good candidate for a binary system. 1FGL J1018.6−5856 has been noted to be positionally coincident with the supernova remnant G284.3−1.8 (*12*) and the TeV source HESS J1018−589 (*13*), although it has not been shown that these sources are actually related.

The modulation at a period of 16.6 days has a power more than 25 times the mean value of the power spectrum, and has a false alarm probability of $3\times10^{-8}$, taking into account the number of statistically independent frequency bins. From both the power spectrum itself (*14*) and from fitting the light curve we derived a period of 16.58 ± 0.02 days. The folded light curve (Fig. 1) has a sharp peak together with additional broader modulation. We modeled this to determine the epoch of maximum flux by fitting a function consisting of the sum of a sine wave and a Gaussian function and obtained $T_{\rm max}$ = MJD 55403.3 ± 0.4.

The gamma-ray spectrum of 1FGL J1018.6−5856 shows substantial curvature through the LAT passband. To facilitate discussion of the lower ($<$ 1 GeV) and higher energy ($>$ 1 GeV) gamma rays, we adopted as our primary model a broken power law with photon indices $\Gamma_{0.1-1}$ and $\Gamma_{1-10}$ for energies below and above 1 GeV respectively. The best-fit values (see SOM) are $\Gamma_{0.1-1} = 2.00 \pm 0.04_{stat} \pm 0.08_{syst}$ and $\Gamma_{1-10} = 3.09 \pm 0.06_{stat} \pm 0.12_{syst}$, along with an integral energy flux above 100 MeV of $(2.8 \pm 0.1_{stat} \pm 0.3_{syst}) \times 10^{-10}$ erg cm$^{-2}$ s$^{-1}$. A power law with exponential cutoff (*7,8*), $dN/dE = N_0(E/\mathrm{GeV})^{-\Gamma}\exp(-E/E_c)$, gives an acceptable fit with $\Gamma = 1.9 \pm 0.1$ and $E_c = 2.5 \pm 0.3$ GeV (statistical errors only). Although this spectral shape is qualitatively similar to that of pulsars and also LS I +61° 303 and LS 5039, so far no detection of pulsed gamma-ray emission has been reported (*15*).

To investigate variability on the 16.6 day period we folded the data into 10 uniform bins in orbital phase, and within each phase bin refit the broken power law parameters. The resulting folded light curve (Fig. 2) indicates substantial variability in both the source brightness and spectral shape. In agreement with the detection of multiple harmonics of the orbital period in the power spectrum, there appear to be two primary features. For phases 0.2–0.6, the spectral curvature decreases and the peak of the spectral energy distribution lies below the LAT passband (indicated by $\Gamma_{0.1-1} > 2$). The onset of this soft spectrum is approximately coincident with a rise in X-ray emission and a peak in radio emission discussed below. A weaker peak appears in the low-energy ($< 1$ GeV) $\gamma$-ray flux at phase 0.5 (Fig. 2). For the remaining phases, the LAT spectrum hardens with a comparatively sharp rise to, and fall from, a peak around 1 GeV ($\Gamma_{0.1-1} < 2$, $\Gamma_{1-10} > 2$). The variable spectral shape implies that only a modest fraction of the flux could be steady, magnetospheric emission from a pulsar.

We undertook observations of the location of 1FGL J1018.6$-$5856 covering the 0.3 – 10 keV energy range using the X-ray Telescope (XRT) onboard the Swift satellite. The first observation was obtained on 29 September 2009 with an exposure of 5 ks. A single source was detected in the XRT image (Fig. 3) within the LAT error circle. We then obtained additional observations from January to April 2011 to search for X-ray variability (see SOM) and found large amplitude variability. Folded on the gamma-ray ephemeris (Fig. 4), there is a sharp peak in X-ray flux, coincident with the gamma-ray peak. However, in addition to this, a sinewave-like periodic modulation is also seen that peaks near phase 0.3 to 0.4.

Swift Ultra-Violet/Optical Telescope (UVOT) (*16*) observations were obtained simultaneously with the X-ray observations. The X-ray source is positionally coincident with a bright source seen in the UVOT images (Fig. 3, SOM) which in turn is coincident with a source in the United States Naval Observatory B1.0 catalog at (J2000.0) R.A. = $10^h\ 18^m\ 55^s.60 \pm 0.1''$, decl. = $-58^\circ\ 56'\ 46.2'' \pm 0.1''$. Spectroscopic observations of the optical counterpart were performed using the South African Astronomical Observatory 1.9m telescope and the 2.5m telescope at the Las Campanas Observatory. Absorption lines due to H, He I and He II identify it as an early type star. We used a spectral atlas (*17*) to estimate the spectral type. He II $\lambda$4686 is present in absorption which indicates a main sequence star. The ratio of He II $\lambda$4541 to He I $\lambda$4471 implies an O6 spectral type. Weak emission is seen from N III but not He II, which indicates an ((f)) classification. We therefore estimate the spectral type as O6V((f)). This is very similar to the spectral type of LS 5039 (*18*). Interstellar absorption bands provide an estimate of the reddening; from the features at 4430 and 5780 Å, we derive $E(B-V)$ = 0.9 and 1.6 respectively. Taking $V$ ~12.6 from measurements with the All Sky Automated Survey (ASAS) (*19*) we derive a distance, $d = 5 \pm 2$ kpc, allowing for uncertainties in the reddening and spectral classification.

Radio observations of the 1FGL J1018.6−5856 region were obtained with the Australia Telescope Compact Array (ATCA) at frequencies of 5.5 and 9 GHz. A faint radio source at R.A. = $10^h$ $18^m$ $55^s$.580, decl. = –58° 56′ 45.5″ (± 0.1″, 0.3″ respectively) is coincident with the stellar position. The radio source was clearly seen to be variable (Fig. 4). Unlike the gamma-ray and X-ray modulation, there is no obvious brightening in the radio at phase zero. Instead it appears that the radio may be following the smoother sine-wave like component of the X-ray modulation.

1FGL J1018.6−5856 shares many properties with LS 5039. They are both fairly steady gamma-ray sources on long timescales, their periodic modulations have not shown large changes, and their optical counterparts are of a very similar spectral type. The X-ray light curve of LS 5039 appears to be highly repeatable (*20,21*), and the X-ray lightcurve of 1FGL J1018.6−5856 also shows repeatable behavior with a flux increase around phase 0 repeated over four orbital periods. The lack of variability in UV/optical brightness is also reminiscent of LS 5039 (*22,23*). This suggests that there is little ellipsoidal modulation of the primary star and hence that it sub-stantially underfills its Roche lobe. On the other hand, the relative phasing of the gamma-ray spectral modulation and flux modulation differ from those of LS 5039 where the spectrum is softest when the flux is highest (*8*). Also, for LS 5039 the phases of maximum X-ray and gamma-ray do not coincide (*8, 21*). The brightest peak in the folded gamma-ray light curve of 1FGL J1018.6−5856 at phase 0, is associated with the hardest gamma-ray spectrum and is coincident with X-ray flaring and minimum radio emission. Finally, 1FGL J1018.6−5856 has a much longer orbital period.

The gamma-ray modulation observed in 1FGL J1018.6−5856 could be due to anisotropic inverse Compton (IC) scattering between stellar photons and high-energy electrons that varies with orbital phase, as proposed for LS 5039 and LS I +61° 303 (*7,8*). However, the modulation amplitude is considerably lower in 1FGL J1018.6−5856 ($(f_{\rm max} - f_{\rm min})/(f_{\rm max} + f_{\rm min}) \approx 25\%$) compared to LS 5039 (≈ 60%). Modulation amplitude should increase with eccentricity, and is highest for systems viewed edge-on (*24*); however, in the case of LS I +61° 303, the modulation fraction has been observed to undergo large changes (*25*). If the IC scattering interpretation is correct, then this implies that 1FGL J1018.6−5856 has both low inclination and low eccentricity. For comparison, the eccentricity of LS 5039 has been reported to be in the range of 0.3 to 0.5 (*18, 26, 27*). Although a low inclination angle implies that it would be difficult to measure the radial velocity of the companion from optical studies, the small Doppler shifts predicted would facilitate a pulsation search at GeV energies.

The gamma-ray spectral variability of 1FGL J1018.6−5856 over the orbit is also reminiscent of LS 5039, but unlike the behavior of LS I +61° 303. If the high energy electron distribution remains constant along the orbit, spectral changes are expected due to the anisotropic IC cross-section only if the inclination is substantial. In this case, harder spectra are expected to occur when the stellar photons are forward-scattered by the electrons (i.e., at inferior conjunction), which is also typically when the scattering rate is at its orbital minimum. However, for 1FGL J1018.6−5856 the hardness ratio and flux are correlated, unlike for LS 5039 (*8*). If periastron passage coincides with inferior conjunction then a high photon density might compensate for the unfavorable interaction angle but this requires fine-tuning. The spectral variability is more likely to reflect intrinsic variations, for instance in the cooling of emitting particles. Moreover, both PSR B1259−63 and LS I +61° 303 (*7, 9*) show that a simple model may not be correct. The phasing of gamma-ray maximum at GeV energies is not consistent with IC scattering on stellar photons, as it is delayed in both PSR B1259−63 and LS I +61° 303, implying other mechanisms may be at work. For example, there could be other seed photon sources, Doppler boosting, or other radiative mechanisms at work.

The gamma-ray energy flux of 1FGL J1018.6−5856 implies a luminosity of $\sim 8\times 10^{35}$ ($d$/5 kpc)$^2$ ergs s$^{-1}$ ($E > 100$ MeV), while the implied X-ray luminosity is highly variable with fluxes up to $\sim 10^{34}$ ($d$/5kpc)$^2$ ergs s$^{-1}$. For comparison, the gamma-ray luminosity of LS 5039 is $\sim 2\times 10^{35}$ ($d$/2.5 kpc)$^2$ ergs s$^{-1}$ (*25*). This is somewhat surprising; compared to LS 5039 the longer orbital period by a factor 4 implies a major axis larger by a factor 2.5 so that the mean stellar radiation density seen by the compact object is smaller by a factor 6. The higher gamma-ray luminosity of 1FGL J1018.6−5856 indicates the power injected in non-thermal particles must therefore be substantially higher in 1FGL J1018.6−5856 than in LS 5039. The similarity with LS 5039 suggests that we may be observing a rapidly rotating neutron star interacting with its companion. This raises the possibility that the neutron star rotation period might be detectable as is the case with PSR B1259−63. However, our observations cannot definitely exclude an accreting neutron star or black hole.

**Acknowledgments:** The Fermi LAT Collaboration acknowledges support from a number of agencies and institutes for both development and the operation of the LAT as well as scientific data analysis. These include NASA and DOE in the United States, CEA/Irfu and IN2P3/CNRS in France, ASI and INFN in Italy, MEXT, KEK, and JAXA in Japan, and the K. A. Wallenberg Foundation, the Swedish Research Council and the National Space Board in Sweden. Additional support from INAF in Italy and CNES in France for science analysis during the operations phase is also gratefully acknowledged. Fermi LAT data are available from the Fermi

Science Support Center: http://fermi.gsfc.nasa.gov/ssc/. This work made use of data supplied by the UK Swift Science Data Centre at the University of Leicester.

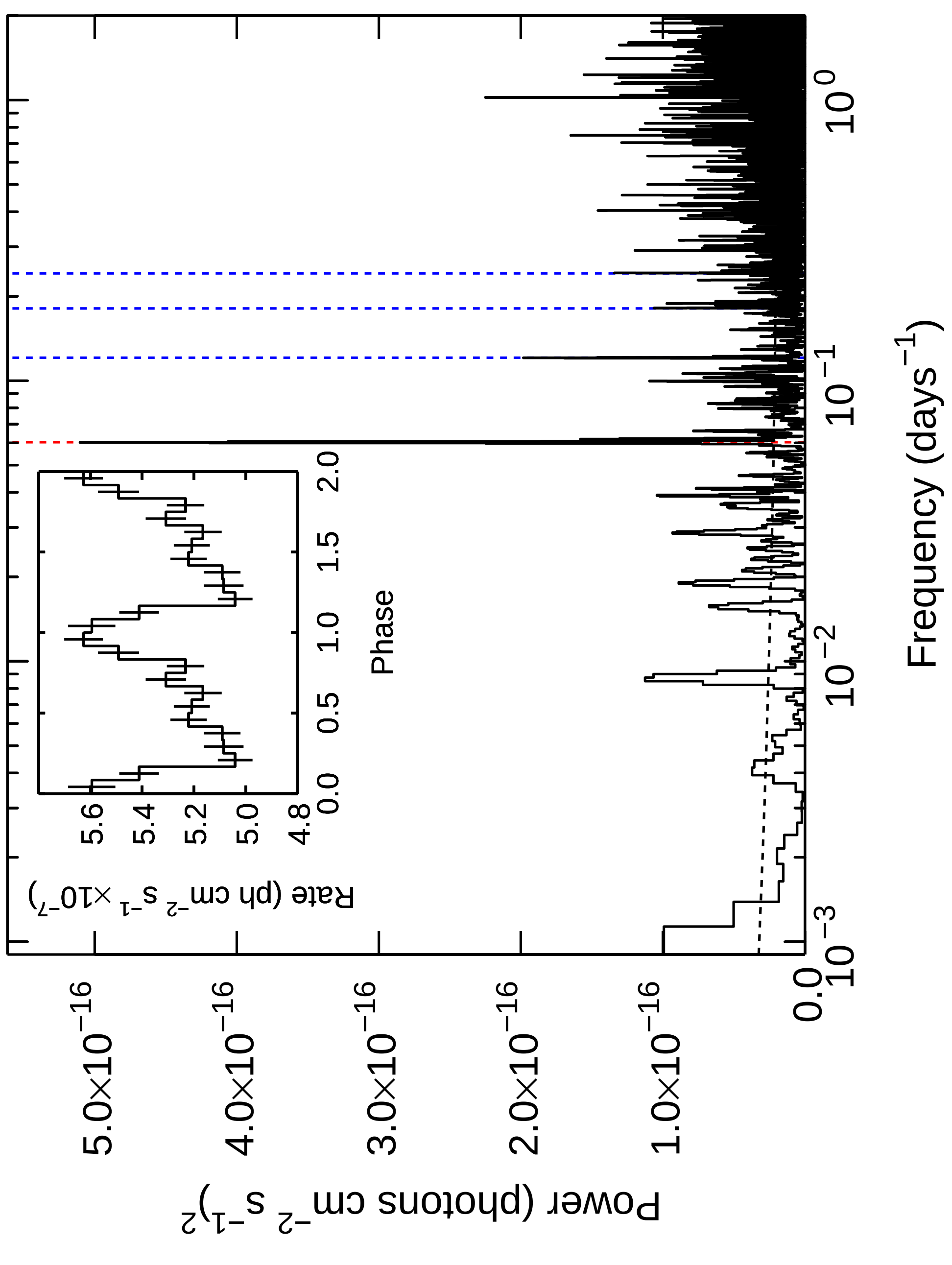

**Fig. 1.** Power spectrum of the LAT weighted photon light curve ($E > 100$ MeV) of 1FGL J1018.6−5856. The power spectrum is oversampled by a factor of 4 compared to its nominal resolution. The red dashed line indicates the 16.6 day period and the blue dashed lines the second, third and fourth harmonics of this. The dashed black line is a fit to the continuum power. The inset shows the weighted photon light curve folded on the 16.6 day period.

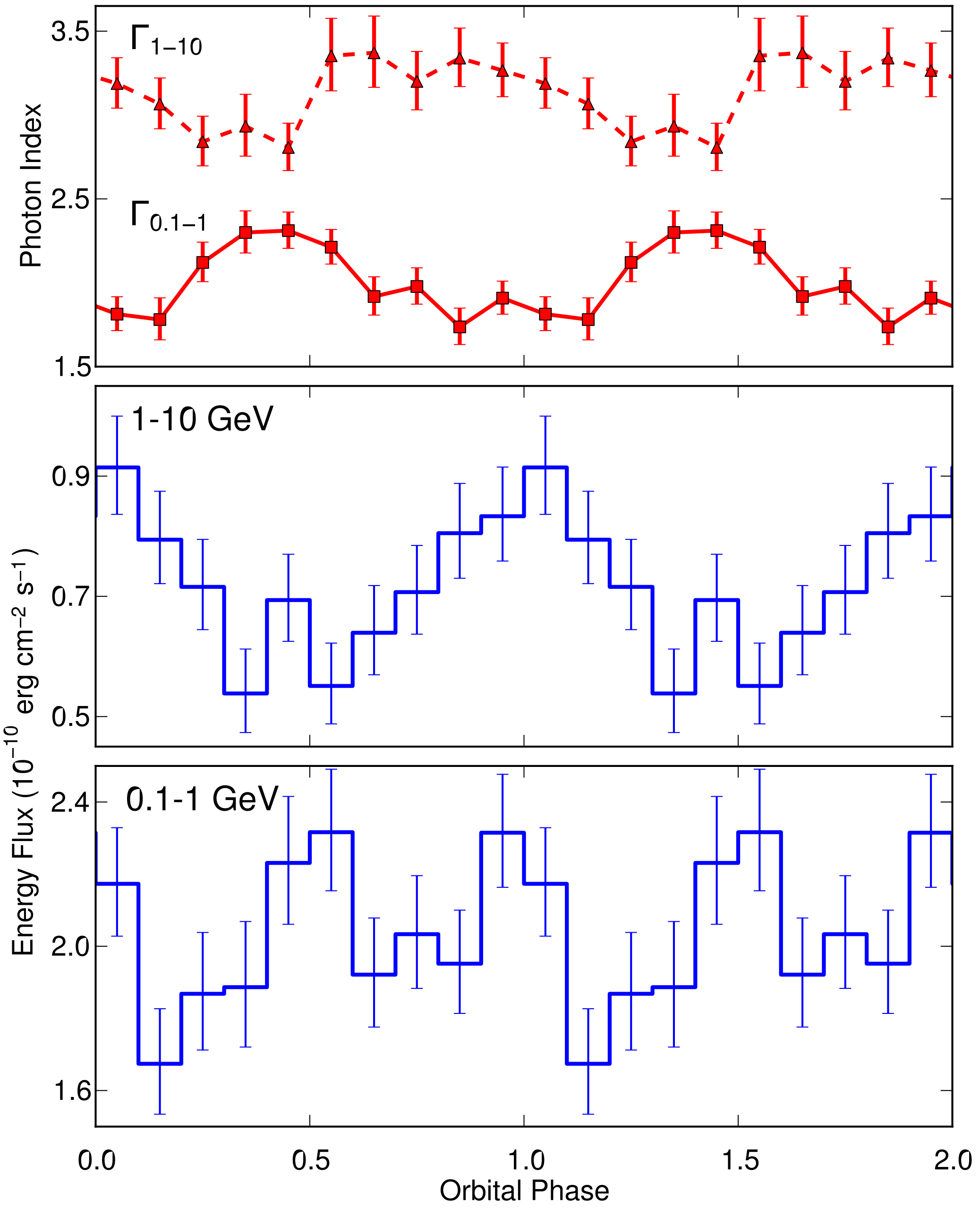

**Fig. 2.** The orbital modulation of the flux and spectral indices of 1FGL J1018.6−5856 in the 0.1 - 10 GeV band as measured with the Fermi LAT. $\Gamma_{0.1-1}$ and $\Gamma_{1-10}$ are photon spectral indices for energies below and above 1 GeV, respectively, using a broken power law model.

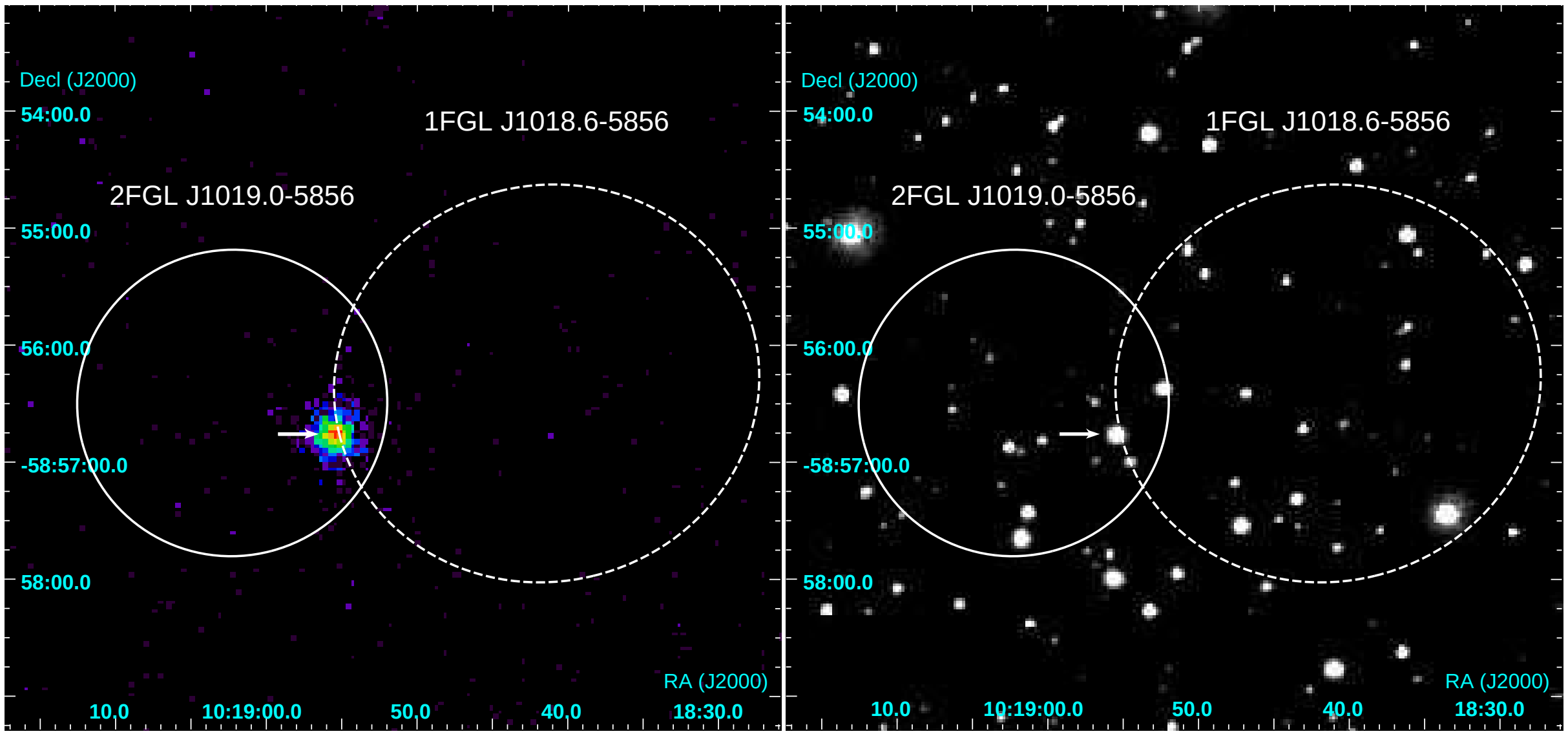

**Fig. 3.** Swift XRT X-ray (left) and UVOT-W1 (right) images of the region around 1FGL J1018.6−5856. The X-ray/optical counterpart is marked by an arrow near the center of both images. The LAT 95% confidence ellipses from the 1FGL (*12*) and 2FGL (*28*) catalogs are marked.

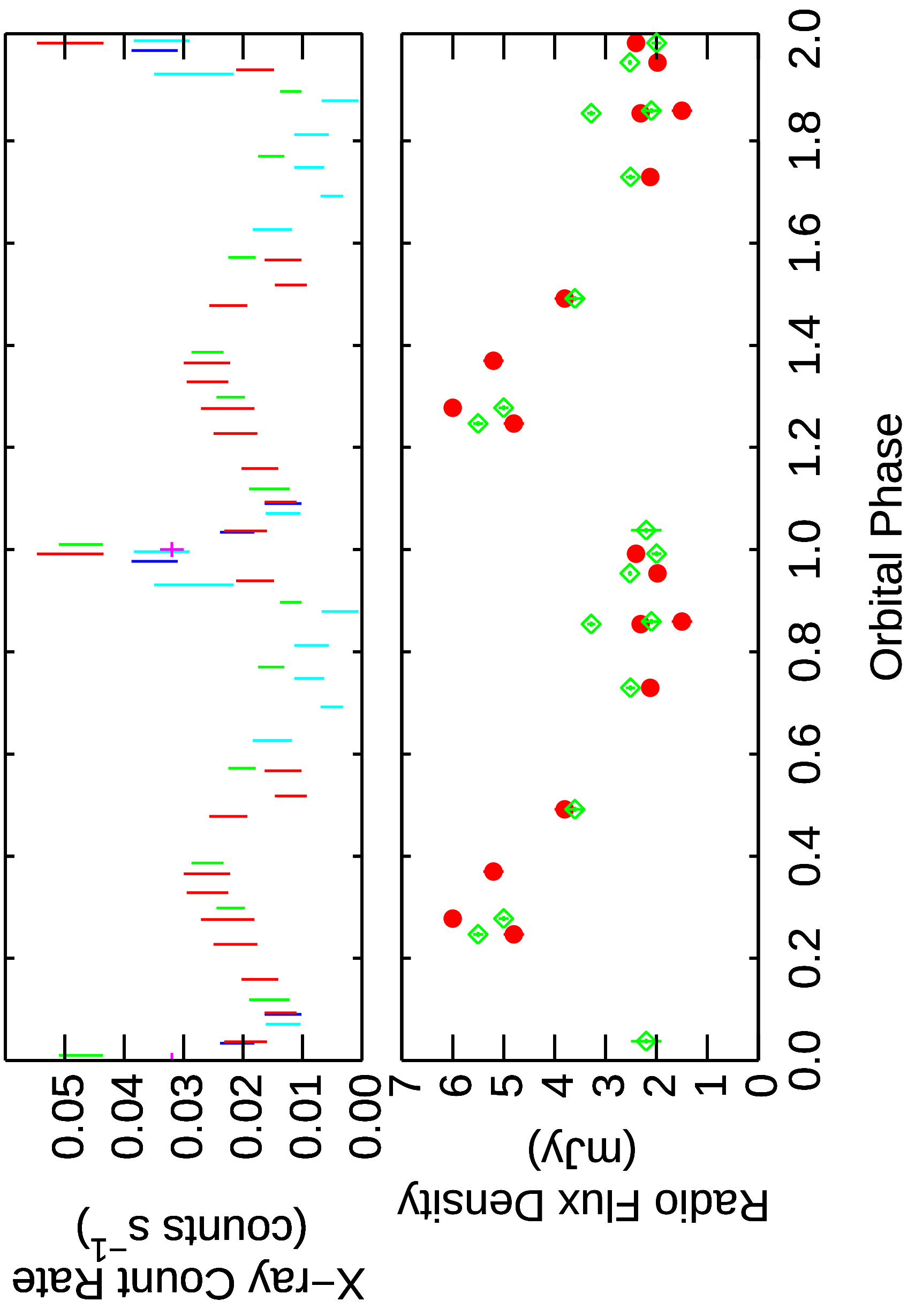

**Fig. 4.** X-ray (upper panel) and radio (lower panel) observations of 1FGL J1018.6−5856 folded on the orbital period. The X-ray data are from the Swift XRT and cover the energy range 0.3 to 10 keV. For the X-ray observations the different colors indicate data taken from different 16.58 day orbital cycles. For the radio data the green diamonds indicate 9 GHz and red circles 5.5 GHz data. The radio data are from the ATCA.

M. Ackermann[1], M. Ajello[1], J. Ballet[2], G. Barbiellini[3,4], D. Bastieri[5,6], A. Belfiore[7,8,9], R. Bellazzini[10], B. Berenji[1], R. D. Blandford[1], E. D. Bloom[1], E. Bonamente[11,12], A. W. Borgland[1], J. Bregeon[10], M. Brigida[13,14], P. Bruel[15], R. Buehler[1], S. Buson[5,6], G. A. Caliandro[16], R. A. Cameron[1], P. A. Caraveo[9], E. Cavazzuti[17], C. Cecchi[11,12], Ö. Çelik[18,19,20], E. Charles[1], S. Chaty[2], A. Chekhtman[21], C. C. Cheung[22*], J. Chiang[1], S. Ciprini[23,12], R. Claus[1], J. Cohen-Tanugi[24], S. Corbel[2,25], R. H. D. Corbet[18,20*], S. Cutini[17], A. de Luca[26], P. R. den Hartog[1], F. de Palma[13,14], C. D. Dermer[27], S. W. Digel[1], E. do Couto e Silva[1], D. Donato[19,28], P. S. Drell[1], A. Drlica-Wagner[1], R. Dubois[1], G. Dubus[29,30], C. Favuzzi[13,14], S. J. Fegan[15], E. C. Ferrara[18], W. B. Focke[1], P. Fortin[15], Y. Fukazawa[31], S. Funk[1], P. Fusco[13,14], F. Gargano[14], D. Gasparrini[17], N. Gehrels[18], S. Germani[11,12], N. Giglietto[13,14], F. Giordano[13,14], M. Giroletti[32], T. Glanzman[1], G. Godfrey[1], I. A. Grenier[2], J. E. Grove[27], S. Guiriec[33], D. Hadasch[16], Y. Hanabata[31], A. K. Harding[18], M. Hayashida[1,34], E. Hays[18], A. B. Hill[35], R. E. Hughes[36], G. Jóhannesson[37], A. S. Johnson[1], T. J. Johnson[22], T. Kamae[1], H. Katagiri[38], J. Kataoka[39], M. Kerr[1*], J. Knödlseder[40,41], M. Kuss[10], J. Lande[1], F. Longo[3,4], F. Loparco[13,14], M. N. Lovellette[27], P. Lubrano[11,12], M. N. Mazziotta[14], J. E. McEnery[18,28], P. F. Michelson[1], W. Mitthumsiri[1], T. Mizuno[31], C. Monte[13,14], M. E. Monzani[1], A. Morselli[42], I. V. Moskalenko[1], S. Murgia[1], T. Nakamori[39], M. Naumann-Godo[2], J. P. Norris[43], E. Nuss[24], M. Ohno[44], T. Ohsugi[45], A. Okumura[1,44], N. Omodei[1], E. Orlando[1,46], M. Ozaki[44], D. Paneque[47,1], D. Parent[48], M. Pesce-Rollins[10], M. Pierbattista[2], F. Piron[24], G. Pivato[6], T. A. Porter[1], S. Rainò[13,14], R. Rando[5,6], M. Razzano[10,7], A. Reimer[49,1], O. Reimer[49,1], S. Ritz[7], R. W. Romani[1], M. Roth[50], P. M. Saz Parkinson[7], C. Sgrò[10], E. J. Siskind[51], G. Spandre[10], P. Spinelli[13,14], D. J. Suson[52], H. Takahashi[45], T. Tanaka[1], J. G. Thayer[1], J. B. Thayer[1], D. J. Thompson[18], L. Tibaldo[5,6], M. Tinivella[10], D. F. Torres[16,53], G. Tosti[11,12], E. Troja[18,54], Y. Uchiyama[1], T. L. Usher[1], J. Vandenbroucke[1], G. Vianello[1,55], V. Vitale[42,56], A. P. Waite[1], B. L. Winer[36], K. S. Wood[27], M. Wood[1], Z. Yang[57,58], S. Zimmer[57,58], M. J. Coe[35], F. Di Mille[59], P. G. Edwards[60], M. D. Filipović[61], J. L. Payne[61], J. Stevens[62], M. A. P. Torres[63]

1. W. W. Hansen Experimental Physics Laboratory, Kavli Institute for Particle Astrophysics and Cosmology, Department of Physics and SLAC National Accelerator Laboratory, Stanford University, Stanford, CA 94305, USA
2. Laboratoire AIM, CEA-IRFU/CNRS/Université Paris Diderot, Service d'Astrophysique, CEA Saclay, 91191 Gif sur Yvette, France
3. Istituto Nazionale di Fisica Nucleare, Sezione di Trieste, I-34127 Trieste, Italy
4. Dipartimento di Fisica, Università di Trieste, I-34127 Trieste, Italy
5. Istituto Nazionale di Fisica Nucleare, Sezione di Padova, I-35131 Padova, Italy
6. Dipartimento di Fisica "G. Galilei", Università di Padova, I-35131 Padova, Italy
7. Santa Cruz Institute for Particle Physics, Department of Physics and Department of Astronomy and Astrophysics, University of California at Santa Cruz, Santa Cruz, CA 95064, USA
8. Università degli Studi di Pavia, 27100 Pavia, Italy
9. INAF-Istituto di Astrofisica Spaziale e Fisica Cosmica, I-20133 Milano, Italy
10. Istituto Nazionale di Fisica Nucleare, Sezione di Pisa, I-56127 Pisa, Italy
11. Istituto Nazionale di Fisica Nucleare, Sezione di Perugia, I-06123 Perugia, Italy
12. Dipartimento di Fisica, Università degli Studi di Perugia, I-06123 Perugia, Italy
13. Dipartimento di Fisica "M. Merlin" dell'Università e del Politecnico di Bari, I-70126 Bari, Italy
14. Istituto Nazionale di Fisica Nucleare, Sezione di Bari, 70126 Bari, Italy

15. Laboratoire Leprince-Ringuet, École polytechnique, CNRS/IN2P3, Palaiseau, France

16. Institut de Ciències de l'Espai (IEEE-CSIC), Campus UAB, 08193 Barcelona, Spain

17. Agenzia Spaziale Italiana (ASI) Science Data Center, I-00044 Frascati (Roma), Italy

18. NASA Goddard Space Flight Center, Greenbelt, MD 20771, USA

19. Center for Research and Exploration in Space Science and Technology (CRESST) and NASA Goddard Space Flight Center, Greenbelt, MD 20771, USA

20. Center for Space Sciences and Technology, University of Maryland Baltimore County, Baltimore, MD 21250, USA

21. Artep Inc., 2922 Excelsior Springs Court, Ellicott City, MD 21042, resident at Naval Research Laboratory, Washington, DC 20375, USA

22. National Research Council Research Associate, National Academy of Sciences, Washington, DC 20001, resident at Naval Research Laboratory, Washington, DC 20375, USA

23. ASI Science Data Center, I-00044 Frascati (Roma), Italy

24. Laboratoire Univers et Particules de Montpellier, Université Montpellier 2, CNRS/IN2P3, Montpellier, France

25. Institut universitaire de France, 75005 Paris, France

26. Istituto Universitario di Studi Superiori (IUSS), I-27100 Pavia, Italy

27. Space Science Division, Naval Research Laboratory, Washington, DC 20375-5352, USA

28. Department of Physics and Department of Astronomy, University of Maryland, College Park, MD 20742, USA

29. Institut de Planétologie et d'Astrophysique de Grenoble, Université Joseph Fourier-Grenoble 1 / CNRS-INSU, UMR 5274, Grenoble, F-38041, France

30. Funded by contract ERC-StG-200911 from the European Community

31. Department of Physical Sciences, Hiroshima University, Higashi-Hiroshima, Hiroshima 739-8526, Japan

32. INAF Istituto di Radioastronomia, 40129 Bologna, Italy

33. Center for Space Plasma and Aeronomic Research (CSPAR), University of Alabama in Huntsville, Huntsville, AL 35899, USA

34. Department of Astronomy, Graduate School of Science, Kyoto University, Sakyo-ku, Kyoto 606-8502, Japan

35. School of Physics and Astronomy, University of Southampton, Highfield, Southampton, SO17 1BJ, UK

36. Department of Physics, Center for Cosmology and Astro-Particle Physics, The Ohio State University, Columbus, OH 43210, USA

37. Science Institute, University of Iceland, IS-107 Reykjavik, Iceland

38. College of Science, Ibaraki University, 2-1-1, Bunkyo, Mito 310-8512, Japan

39. Research Institute for Science and Engineering, Waseda University, 3-4-1, Okubo, Shinjuku, Tokyo 169-8555, Japan

40. CNRS, IRAP, F-31028 Toulouse cedex 4, France

41. GAHEC, Université de Toulouse, UPS-OMP, IRAP, Toulouse, France

42. Istituto Nazionale di Fisica Nucleare, Sezione di Roma “Tor Vergata”, I-00133 Roma, Italy

43. Department of Physics, Boise State University, Boise, ID 83725, USA

44. Institute of Space and Astronautical Science, JAXA, 3-1-1 Yoshinodai, Chuo-ku, Sagamihara, Kanagawa 252-5210, Japan

45. Hiroshima Astrophysical Science Center, Hiroshima University, Higashi-Hiroshima, Hiroshima 739-8526, Japan

46. Max-Planck Institut für extraterrestrische Physik, 85748 Garching, Germany

47. Max-Planck-Institut für Physik, D-80805 München, Germany

48. Center for Earth Observing and Space Research, College of Science, George Mason University, Fairfax, VA 22030, resident at Naval Research Laboratory, Washington, DC 20375, USA

49. Institut für Astro- und Teilchenphysik and Institut für Theoretische Physik, Leopold-Franzens-Universität Innsbruck, A-6020 Innsbruck, Austria

50. Department of Physics, University of Washington, Seattle, WA 98195-1560, USA

51. NYCB Real-Time Computing Inc., Lattingtown, NY 11560-1025, USA

52. Department of Chemistry and Physics, Purdue University Calumet, Hammond, IN 46323-2094, USA

53. Institució Catalana de Recerca i Estudis Avançats (ICREA), Barcelona, Spain

54. NASA Postdoctoral Program Fellow, USA

55. Consorzio Interuniversitario per la Fisica Spaziale (CIFS), I-10133 Torino, Italy

56. Dipartimento di Fisica, Università di Roma "Tor Vergata", I-00133 Roma, Italy

57. Department of Physics, Stockholm University, AlbaNova, SE-106 91 Stockholm, Sweden

58. The Oskar Klein Centre for Cosmoparticle Physics, AlbaNova, SE-106 91 Stockholm, Sweden

59. Australian Astronomical Observatory - Las Campanas Observatory, Colina, El Pino Casilla 601, La Serena, Chile

60. Australia Telescope National Facility, CSIRO Astronomy and Space Science, Narrabri NSW 2390, Australia

61. University of Western Sydney, Locked Bag 1797, Penrith NSW 2751, Australia

62. CSIRO Astronomy and Space Science, Epping, NSW 1710, Australia

63. Harvard-Smithsonian Center for Astrophysics, Cambridge, MA 02138, USA

* To whom correspondence should be addressed. Email: Robin.Corbet@nasa.gov (R. H. D. Corbet); kerrm@stanford.edu (M. Kerr); Teddy.Cheung.ctr@nrl.navy.mil (C. C. Cheung)

# Supporting Online Material (SOM)

## Fermi-LAT Data Analysis

The Fermi LAT is a pair conversion telescope designed to cover the energy band from 20 MeV to greater than 300 GeV (*11*). Fermi operates predominately in a sky-survey mode where the entire sky is observed every ∼3 hours. Analysis was performed using Version 9, Release 18 of the Fermi Science Tools[1], Pass 6 "DIFFUSE" class events, and the P6_V3_DIFFUSE instrument response function (IRF). The LAT data set used here covers the interval from MJD 54,682 to 55,669 (4 August 2008 to 18 April 2011). This is somewhat longer than the data set used in an initial analysis (*29*). In order to maximize signal-to-noise while allowing the use of short time bins we employed a weighted photon technique to extract the light curves that were used in the period searches. This method is similar to aperture photometry. However, the probability that a photon came from the source of interest is calculated and *probabilities* are summed, rather than *photons*. This approach builds on previous work (*30,31*) and has been successfully applied to increase the sensitivity for Fermi pulsar searches (*32*). The probability that a photon came from the source of interest was calculated using `gtsrcprob` based on the fluxes and spectral models of the first Fermi LAT Catalog (*12*).

Power spectra were calculated for all sources in the 1FGL catalog in order to search for sources that displayed periodic modulation and so would be candidate gamma-ray binaries. Because time bins have large variation in exposure we weighted each time bin's contribution to the power spectrum by its relative exposure (*4*). For all power spectra the height of the strongest peak relative to the mean power level was calculated. In addition, all power spectra were visually inspected to enable the identification of sources where power was concentrated at a single frequency, indicating a possible binary, rather than broad band modulation such as exhibited by active galactic nuclei.

[1] http://fermi.gsfc.nasa.gov/ssc/data/analysis

We note that, due to the strong energy dependence of both the LAT point-spread function and the diffuse Galactic background, high energy photons are weighted appreciably more heavily than low energy photons. In light curves, this spectral dependence emphasizes hard features. Accordingly, the weighted-photon light curve for 1FGL J1018.6−5856 is dominated by a hard peak at phase 0, while the soft feature at phase 0.5 is only revealed by likelihood analysis.

To investigate whether the 16.6 day period could be due to some type of systematic effect we investigated the power spectra of two gamma-ray emitting pulsars both located about 1.4° from 1FGL J1018.6−5856: PSR J1023−5746 and PSR J1028−5819 (*12*). Neither showed any modulation on the 16.6 day period. We also employed a different period searching technique that uses photons accumulated in an annulus around a source to more accurately model exposure variations, and uses the $Z^2_m$ method (*33*). This technique again showed highly significant modulation at 16.6 days.

As a third test of the presence of modulation in the LAT light curve we extracted a light curve using maximum likelihood fitting. We divided the data into 2.0 day sections using a 10 degree radius around 1FGL J1018.6−5856, and performed fits. A power-law spectrum was used for 1FGL J1018.6−5856 with the flux and power-law index allowed to vary. Other sources in the region had their parameters fixed at their cataloged values. Only fits that gave test statistic (TS) values greater than 2 were used. We calculated the power spectrum of this light curve for periods longer than 4 days, weighting each data point's contribution by the uncertainty on its flux value. This power spectrum again shows a peak at 16.6 days, at approximately 14 times the mean power level, for a false alarm probability of $\sim 2\times10^{-4}$, allowing for the number of frequencies searched. The reduced statistical significance relative to the other two tests is to be expected, as the TS cut removes some low-flux intervals, there are too few photons in a given interval to constrain the parameters, and 2 days is somewhat long compared to the modulation timescale.

We investigated the long-term stability of the modulation of the gamma-ray flux by dividing the total light curve into four sections and calculating power spectra separately for each interval. We also quantified the modulation during each of these sections by fitting sine waves. Neither procedure showed any significant change in period length or modulation amplitude. We note that the power spectrum of the entire light curve does not exhibit significant low-frequency noise – indicating that 1FGL J1018.6−5856 is stable on long timescales.

The statistical significance of the presence of harmonics in the power spectrum was calculated using "single trial" calculations of the false alarm probability (FAP). From this procedure we find: 2nd harmonic, FAP = $10^{-4}$; 3rd harmonic, FAP = 0.007; 4th harmonic, FAP = 0.001. However, the presence of the 4th harmonic would require a somewhat shorter orbital period of 16.55 days. Because the modulation is non-sinusoidal, a fit of just a sine wave to the periodic modulation does not give a good determination of the time of maximum flux in the weighted photon light curve. Instead, experimentation showed that a good fit to the folded light curve could be obtained with the sum of a sine wave (with period fixed to the orbital period) plus a Gaussian function with a "sigma" of $0.1 \pm 0.03$ of an orbital period. We adopt the time of maximum of the Gaussian component (MJD $55303.3 \pm 0.4$) as phase zero throughout. Fig. 2 shows that this phase zero determined from the weighted photon light curve is consistent with gamma-ray flux maxima in the 0.1 – 1 and 1 – 10 GeV energy bands as well as the X-ray maximum (Fig. 4). We previously reported an epoch of phase zero determined from a sine wave only fit (*29*). This contained a numerical error which fortuitously gave a time of maximum flux consistent with the result from the more complex model.

We performed spectral analysis using the `pointlike` (*32*) tool with cross-checks using the standard `gtlike` tool[2]. In addition to the data selection outlined above, we removed events recorded $> 100^\circ$ from the zenith and when the region of interest impinged too closely on the earth's limb, and we also excluded periods when the spacecraft varied from its typical survey profile. To model background point sources, we used a preliminary version of the 2FGL catalog. We note that 1FGL J1018.6−5856 lies close on the sky to the pulsar PSR J1016−5857 (*34*), but the LAT spatially resolves these two sources. To model diffuse emission we used the *gll_iem_v02* and *isotropic_iem_v02* models of the 1FGL catalog (*12*). In the orbital phase resolved spectral analysis we fixed the background model to the best-fit, phase-averaged values and within each phase bin refit only the parameters of the broken power law. The broken power law model used to fit the spectrum of 1FGL J1018.6−5856 (Fig. S1) is favored over a simple power law with extremely high significance, resulting in an increase in the test statistic (*35*) of 197, or approximately $14\sigma$. We note the broken power law provides a TS increase of 20 over the exponential cutoff model. We assessed systematic errors by repeating our fits with different sky models, with the P6_V11_DIFFUSE IRF, which has been updated to better characterize the instrument's point-spread function and effective area, and with "Pass 7" data, which uses an improved set of algorithms for reconstructing photon events. All of these configurations are consistent within the quoted systematic errors. Finally, we verified that the overall behavior of the energy flux as a function of orbital phase was independent of the spectral model we chose for 1FGL J1018.6−5856.

[2] http://fermi.gsfc.nasa.gov/ssc/data/analysis

# X-ray Observations and Analysis

In an initial 5 ks Swift X-ray Telescope (XRT (*36*)) observation of the location of 1FGL J1018.6−5856 on 29 September 2009 (ID 90191), a single X-ray source was detected within the LAT error circle (Fig. 3). This prompted a new observing campaign (ID 31912) with Swift beginning with six 3-5 ks from 14 – 29 January 2011 (∼1 orbital period), which revealed significant X-ray variability in the source. Folded on the gamma-ray ephemeris, a sharp peak in X-ray flux co-incident with the gamma-ray peak was found. Three further daily 2-3 ks Swift observations (25-27 February) around the next predicted maximum confirmed the X-ray peak ~16.6 days later. To cover a full orbital period, 20 daily (predominantly 2-3 ks) observations from 25 February – 16 March were obtained. These observations confirmed two of the next predicted peaks and delineated a smoother periodic modulation peaking near phase 0.3 to 0.4. A final 10 ks exposure was obtained 17 April to increase photon statistics on the main peak. From the 90 ks cumulative exposure of these 30 monitoring observations from program ID 31912, the best-fit XRT position enhanced by UVOT field astrometry (*37,38*) was (J2000) R.A. = $10^h$ $18^m$ $55^s.71$, decl. = $-58°$ $56'$ $47.2''$ (90% confidence radius = $1.9''$). This is consistent with the more precise position obtained with Chandra (*39*).

For spectral analysis around the X-ray peak, we combined all of the XRT exposures within phase = $0.0 \pm 0.05$. The data were best fit with an absorbed single power-law with photon index, $\Gamma = 1.26 \pm 0.25$, absorption, $n_{\rm H} = (0.50 \pm 0.24) \times 10^{22}$ cm$^{-2}$, and 0.3–10 keV observed flux = $2.6\ (+0.3/-0.6) \times 10^{-12}$ ergs cm$^{-2}$ s$^{-1}$ (reduced $\chi^2 = 1.03$ for 27 degrees of freedom). The photon index and absorption around phase 0 are consistent with those derived from Chandra and XMM observations around phases ~0.31-0.32 and ~0.64-0.65, respectively (*39*). The (unabsorbed) 0.3–10 keV luminosity for these 3 phase periods varies between ~(4 – 10) $\times 10^{33}$ $(d/5{\rm kpc})^2$ ergs s$^{-1}$.

# Optical Observations and Analysis

The Swift UVOT observations of the optical counterpart of 1FGL J1018.6$-$5856 yield average magnitudes of $U = 13.34 \pm 0.02$ (3465Å), $W1 = 14.40 \pm 0.03$ (2600Å), $W2 = 15.44 \pm 0.04$ (1928Å), and $M2 = 16.05 \pm 0.02$ (2246Å). There is no notable change in the brightness from the averages (<0.02 mag difference) in the observations. The USNO B1.0 catalog gives magnitudes for this source of $B2 = 13.1$, $R2 = 12.4$, and $I = 11.1$, that have typical uncertainties of 0.3 mag (*40*). Additionally, this optical source is coincident with 2MASS J10185560$-$585645 (*41*) with near-infrared magnitudes of $J = 10.44$, $H = 10.14$, and $K_{\rm s} = 10.02$ (uncertainties of 0.02 mag).

Observations of the optical candidate were performed using the South African Astronomical Observatory (SAAO) 1.9m telescope on 7 February 2011. The Grating Spectrograph with SITe CCD was employed with the #7 grating. The spectrum covers approximately 3600 to 7550 Å with a resolution of 5Å. Data reduction was performed using Figaro (*42*). The spectrum is shown in Fig. S2 and described in the main text.

The optical counterpart was also observed with the 2.5m telescope at the Las Campanas Observatory (LCO) using the Boller and Chivens spectrograph covering the 3750 – 6900 Å range on 5 and 6 February 2011. A 600 l mm$^{-1}$ grating was used that provided a resolution of 3Å. Data were analyzed using IRAF (*43*). The features from the SAAO spectrum were confirmed in the LCO spectrum (Fig. S3).

Photometric V band observations were extracted from the All Sky Automatic Survey (ASAS) (*19*) data base. Approximately 604 observations were obtained between 17 February 2001 to 1 December 2009. We searched for modulation in these observations at the orbital period of 1FGL J1018.6$-$5856 and no significant modulation was detected.

# Radio Observations and Analysis

The Australia Telescope Compact Array (ATCA) observed 1FGL J1018.6$-$5856 on ten occasions between 7 February and 4 May 2011. Observations were made simultaneously at 5.5 GHz and 9.0 GHz, with bandwidths of 2 GHz centered on these frequencies provided by the Compact Array Broadband Backend (*44*). Instrumental issues affected two epochs, resulting in only one frequency yielding useful results in each. Observations were typically made over a 2 – 3 hour period, and the six-element Compact Array was in several different array config-urations over the three month period. The radio position for 1FGL J1018.6$-$5856 reported in the main text was determined from an observation in the most extended array configuration, which provides the highest angular resolution. Where possible, PKS B1934$-$638, the ATCA primary flux-density calibrator, was used for flux-density calibration. For those observations made when PKS B1934$-$638 was below the telescope's horizon, the secondary calibrator PKS B0823$-$500 was used. The flux density of PKS B0823$-$500 is known to vary slowly with time, and its flux density was calibrated against PKS B1934$-$638 within a week for the epochs it was used. The presence of another source in the field at 5.5 GHz, (at R.A. = $10^h$ $18^m$ $55^s$, decl. = $-58°$ $59'$ $50''$, J2000) enabled a check of the flux-density calibration to be made as this source showed no evidence for significant variability over the first nine epochs, being 2.1 $\pm$ 0.1 mJy beam$^{-1}$. A further consistency check was made using the flux density of the phase calibrator PMN J1047$-$6217. Although this calibrator had a variable flux density, its spectral index remained constant over the period of observations. The positional error obtained from the ATCA observations is noticeably worse in declination, as a several hour observation with an east-west array results in an elongated, or elliptical, beam, with poorer resolution in one direction. Our positional errors also take into account a $\sim$1 milli-arcsec uncertainty on the location of the phase calibration source, but are dominated by the position uncertainty of 1FGL J1018.6$-$5856 itself.

The radio spectral index is clearly variable (Fig. S4), including changing from positive to negative, possibly because of varying absorption in the stellar wind. However, as the observations were made over approximately 5 orbital periods, it is not possible to disentangle variations within one cycle with longer term variations and so we cannot yet identify any orbital phase dependence in the index variations. A physical interpretation is complicated by this ambiguity, and additional radio observations are required to resolve this. We note, however, that the radio behavior of 1FGL J1018.6$-$5856 is different from that of LS 5039 where there is no strong variation of flux or spectral index with orbital phase (*45*).

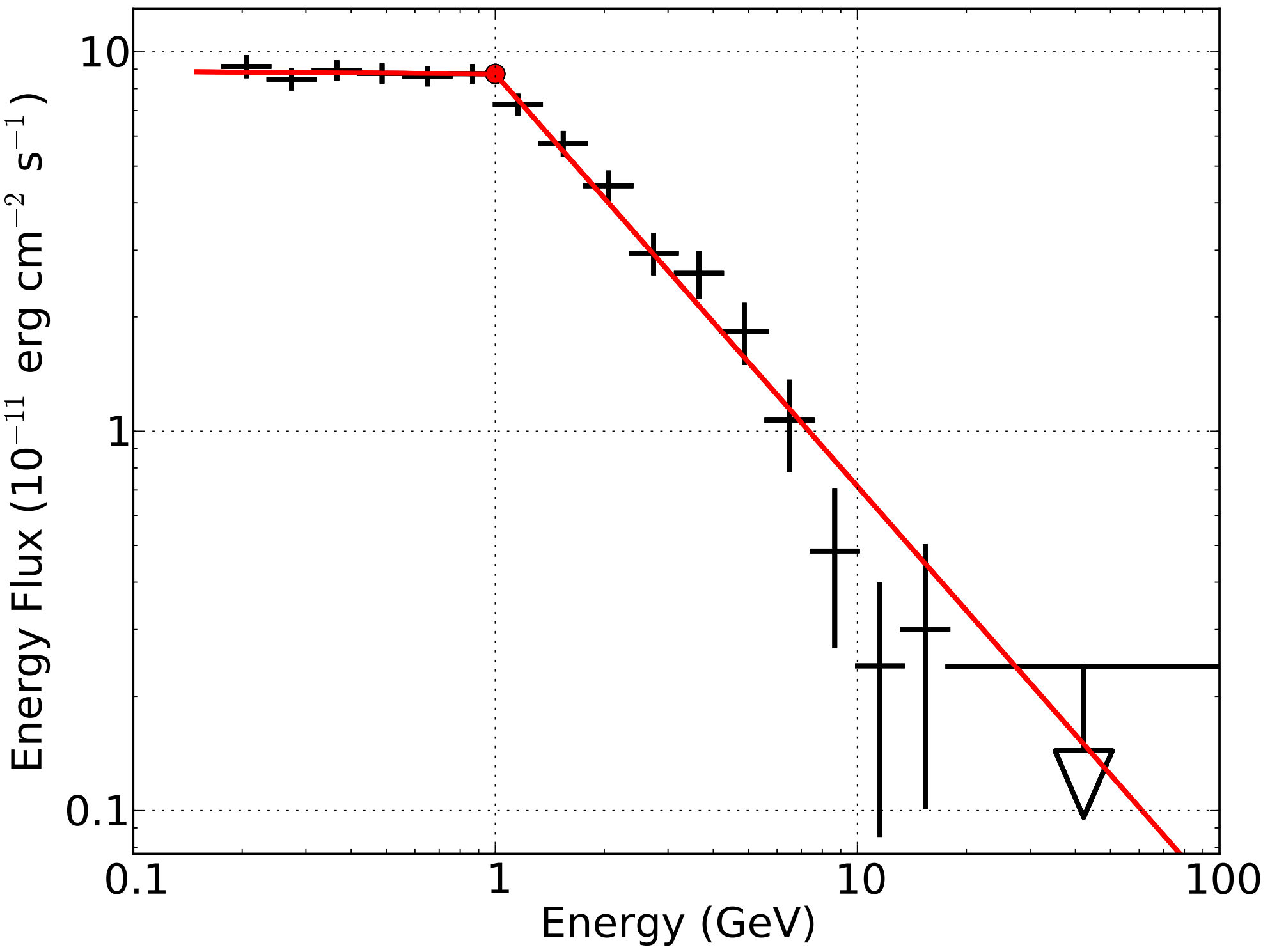

**Fig. S1.** The gamma-ray spectrum of 1FGL J1018.6−5856 obtained with the Fermi-LAT (black error bars). The red line shows the broken power-law model described in the text.

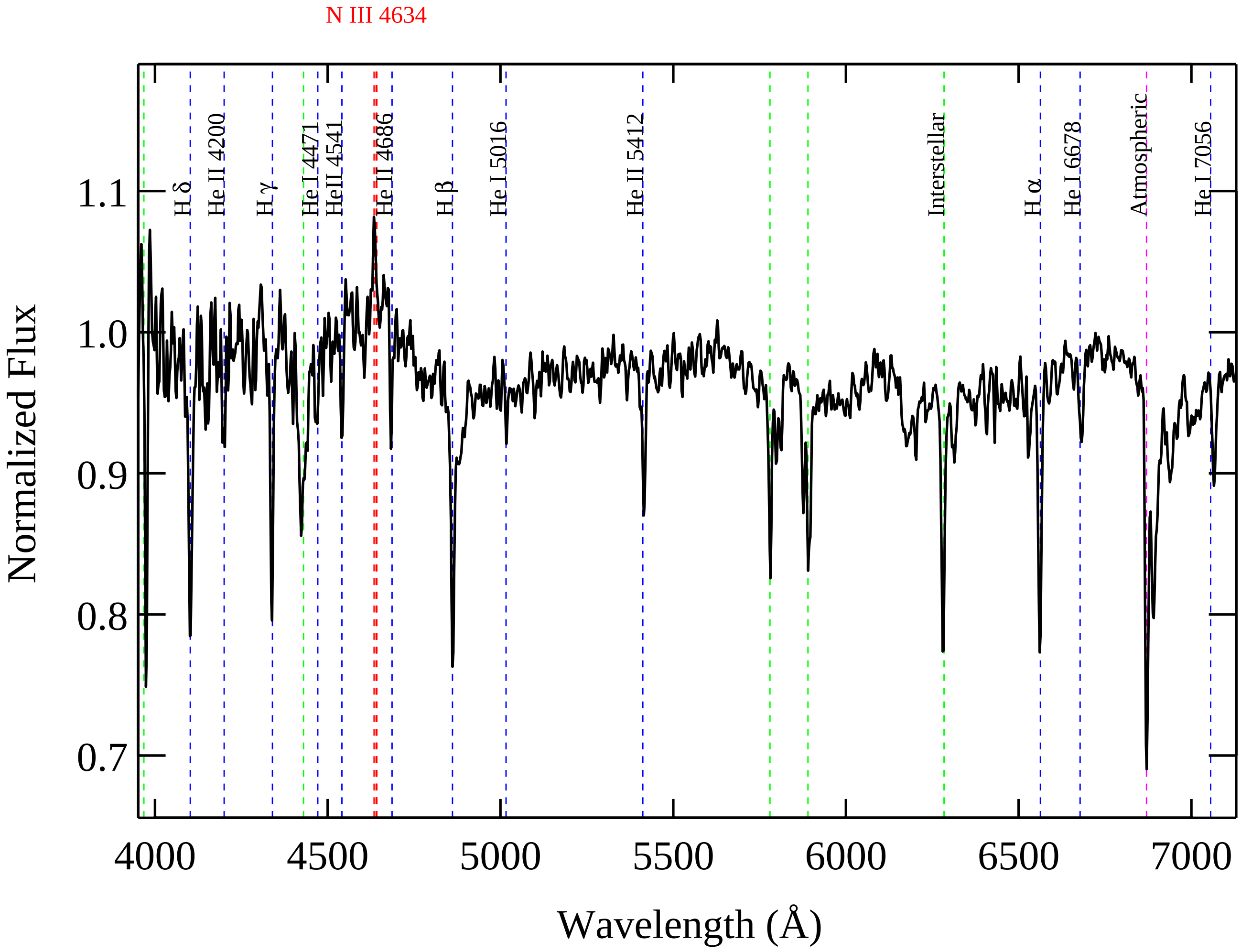

**Fig. S2.** Optical spectrum of the counterpart of 1FGL J1018.6−5856 obtained with the SAAO 1.9m telescope. Instrumental response has been approximately removed by the subtraction of a 5th-order polynomial. Spectral line identifications are marked. The color coding is: blue = stellar absorption, red = stellar emission, green = interstellar absorption, magenta = atmospheric.

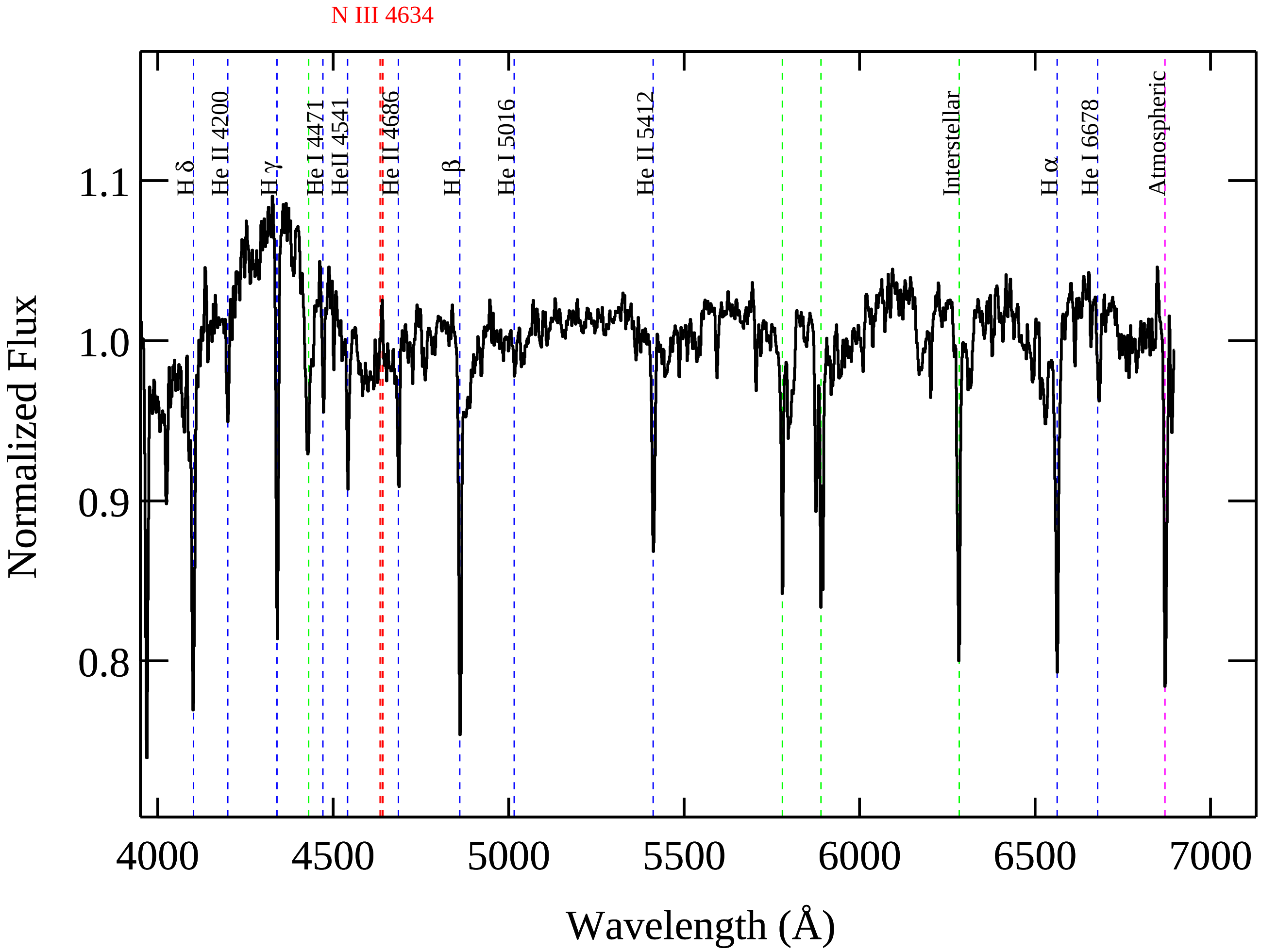

**Fig. S3.** Optical spectrum of the counterpart of 1FGL J1018.6−5856 obtained with the 2.5m telescope at the Las Campanas Observatory (LCO). Instrumental response has been approximately removed by the subtraction of a 9th-order polynomial. Spectral line identifications are marked. The color coding is: blue = stellar absorption, red = stellar emission, green = interstellar absorption, magenta = atmospheric.

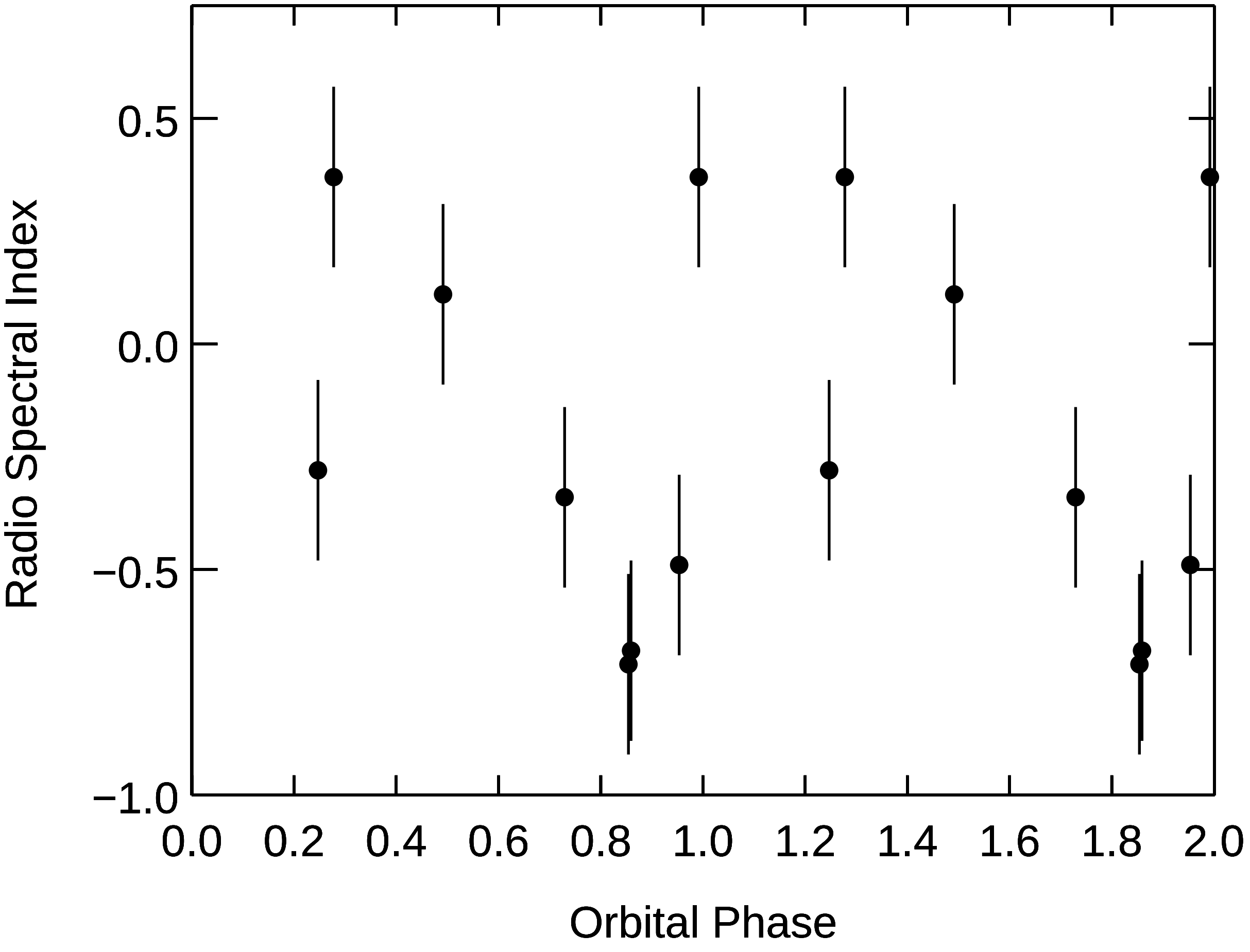

**Fig. S4.** The radio spectral index of 1FGL J1018.6−5856 derived from ATCA observations at 5.5 and 9 GHz. A conservative estimate of the uncertainty in the radio spectral index, incorporating both statistical and systematic errors, of 0.2 is adopted for all measurements. The radio spectral index ($\alpha$) is defined by $S \propto \nu^{-\alpha}$, where $S$ is the flux density and $\nu$ is the observation frequency.